\documentclass[seceq,supplement]{ptptex}

\usepackage{graphicx}
\usepackage{wrapft}
\usepackage{bm}

\def\la{\mathrel{\mathpalette\fun <}}
\def\ga{\mathrel{\mathpalette\fun >}}
\def\fun#1#2{\lower3.6pt\vbox{\baselineskip0pt\lineskip.9pt
\ialign{$\mathsurround=0pt#1\hfil##\hfil$\crcr#2\crcr\sim\crcr}}}

\def\ve{\varepsilon}

\newcommand{\beq}{\begin{equation}}
\newcommand{\eeq}{\end{equation}}
\newcommand{\bea}{\begin{eqnarray}}
\newcommand{\eea}{\end{eqnarray}}

\newcommand{\bfi}[1]{\mbox{\boldmath $#1$}}

\newcommand{\vb}{{\bfi b}}
\newcommand{\vk}{{\bfi k}}

\newcommand{\vp}{{\bfi p}}
\newcommand{\vq}{{\bfi q}}

\newcommand{\vrr}{{\bfi r}}
\newcommand{\vR}{{\bfi R}}

\newcommand{\vy}{{\bfi y}}

\title{
Recent development of CDCC
}

\author{Masanobu Yahiro, Takuma Matsumoto, Kosho Minomo, \\
 Takenori Sumi and Shin Watanabe}

\inst{Department of Physics, Kyushu University, Fukuoka 812-8581, Japan}

\recdate{\today}%            %Editorial Office will fill in this.

\abst{
This paper shows a brief review on CDCC and 
the microscopic reaction theory as a fundamental theory of CDCC.  
The Kerman-McManus-Thaler theory for nucleon-nucleus scattering 
is extended to nucleus-nucleus scattering.
New development of four-body CDCC is presented. 
An accurate method of treating inclusive reactions is presented 
as an extension of CDCC and the Glauber model. 
}

\begin{document}

\maketitle

\section{Introduction}

The construction of microscopic reaction theory is one 
of the most important subjects in nuclear physics. 
It is a goal of the nuclear reaction theory. 
Furthermore, the construction is essential for many applications. 
Particularly for the scattering of unstable nuclei, there is no 
reliable phenomenological optical potential, since measurements of the elastic 
scattering are not easy. 
An important theoretical tool of analyzing inclusive reactions 
is the Glauber model~\cite{Glauber}. 
The theoretical foundation of the model is shown 
in Ref.~\citen{Yahiro-Glauber}.
The model is based on the eikonal and the adiabatic approximation. 
It is well known that the adiabatic approximation makes
the removal cross section diverge when 
the Coulomb interaction is included. 
The Glauber model has thus been applied mainly for
lighter targets in which the Coulomb interaction is
negligible; see for example 
Refs.~\citen{Hussein,Hencken,Gade,Ogawa01,Tostevin,Bertulani-92,Bertulani-04}
and Refs.~\citen{Ibrahim,Capel-08} for Coulomb corrections to the Glauber model. 

Meanwhile, the method of continuum discretized coupled channels
(CDCC)~\cite{CDCC-review1,CDCC-review2}
is an accurate method of treating
exclusive reactions such as the elastic scattering and
the elastic breakup reaction in which
the target is not excited. 
The theoretical foundation of CDCC is shown in 
Refs.~\citen{CDCC-foundation1,CDCC-foundation2,CDCC-foundation3}. 
Actually, CDCC has succeeded in reproducing data on the scattering of 
not only stable nuclei but also unstable nuclei; 
see for example Refs.~\citen{Tostevin2,Davids,Eikonal-CDCC,Matsumoto,Egami,
Matsumoto3,Matsumoto4,LSCSM1,LSCSM2,THO-CDCC,4body-CDCC-bin,
Matsumoto:2010mi} and references therein. 
The dynamical eikonal approximation~\cite{DEA}
is also an accurate method of 
treating exclusive reactions at intermediate and high incident energies 
where the eikonal approximation is reliable. 
The nucleon removal reaction is composed of the exclusive
elastic-breakup component and 
the inclusive nucleon-stripping component. 
CDCC and the dynamical eikonal approximation 
can evaluate the elastic-breakup cross section, 
but not the stripping cross section.

The experimental exploration of halo nuclei
is moving from lighter nuclei such
as He and C isotopes to relatively heavier nuclei such as Ne isotopes.
Very recently, Takechi {\it et al.} measured the interaction cross section 
$\sigma_{\rm I}$ for the scattering of $^{28-32}$Ne at 240~MeV/nucleon 
and found that $\sigma_{\rm I}$ is quite large particularly 
for $^{31}$Ne~\cite{Takechi}. 
A halo structure of $^{31}$Ne was reported with the 
experiment on the one-neutron removal reaction~\cite{Nakamura}.
This is the heaviest halo nucleus in the present
stage suggested experimentally and resides in the 
''Island-of-inversion" region. Determining the spin-parity 
of $^{31}$Ne is essential to understand the nature 
of ``Island of inversion".

This paper shows a brief review on recent development of CDCC and 
the microscopic reaction theory that yields the foundation of CDCC.  
We present the microscopic reaction theory in 
Sec.~\ref{Microscopic-reaction-theory} and 
new development of four-body CDCC in Sec.~\ref{Four-body-CDCC}. 
We finally propose an accurate method of treating inclusive reactions 
as an extension of CDCC and the Glauber model in 
Sec.~\ref{Eikonal-reaction-theory}.

\section{Microscopic reaction theory}
\label{Microscopic-reaction-theory}

In this section, we present a microscopic reaction theory for 
nucleus-nucleus scattering. This is an extension of 
the Kerman-McManus-Thaler formalism~\cite{KMT} 
of the multiple scattering theory~\cite{Watson} 
for nucleon-nucleus scattering to nucleus-nucleus scattering. 
In principle this reaction theory is applicable for many cases, 
but we consider the simple case in which the projectile breakup is weak, 
because it is not easy to perform fully-microscopic calculations including 
the projectile breakup. 
In the case, the theory is reduced 
to the double-folding model with the effective nucleon-nucleon 
(NN) interaction. The double-folding model is applied to nucleus-nucleus scattering at intermediate energies, 
particularly the scattering of Ne isotopes. 
The breakup effect is estimated by reducing the microscopic model to 
a three-body model and solving the three-body model with CDCC. 
This section is a brief review of 
Refs.~\citen{Minomo-DWS,Minomo:2011bb,Sumi:2012fr}.

\subsection{Model building}
Let us consider the scattering of projectile (P) on target (T). 
The most fundamental equation for this case is 
the many-body Schr\"odinger equation with the realistic nucleon-nucleon (NN) 
interaction $v_{ij}$. 
The multiple scattering theory~\cite{Watson, KMT} 
for nucleon-nucleus scattering 
was extended to nucleus-nucleus scattering~\cite{Yahiro-Glauber}. 
According to the theory, the many-body Schr\"odinger equation is 
approximated into 
\bea
(K+h_{\rm P}+h_{\rm T}+ \sum_{i \in {\rm P}, j \in {\rm T}} \tau_{ij}-E){\hat \Psi}^{(+)}=0  \;,
\label{schrodinger-effective}
\eea
where
$E$ is an energy of the total system,
$K$ is a kinetic energy of the relative motion between P and T, and
$h_{\rm P}$ ($h_{\rm T}$) is an internal Hamiltonian of P (T). 
Here $\tau_{ij}$ is the effective NN interaction in nuclear medium. 
The Brueckner $g$-matrix has commonly been used as $\tau$ 
in many applications;  see for example 
Refs.~\citen{M3Y,JLM,Brieva-Rook,Satchler-1979,Satchler,CEG,
Rikus-von-Geramb,Amos,CEG07}.
The $g$-matrix interaction includes nuclear-medium effects, 
but not the effect of collective excitations induced by 
surface vibration and rotation of finite nucleus, 
since the interaction is evaluated in nuclear matter. 
The effect of collective excitations is small for intermediate energy 
scattering, as shown later. 

\begin{figure}[htb]
\includegraphics[width=6.5 cm]{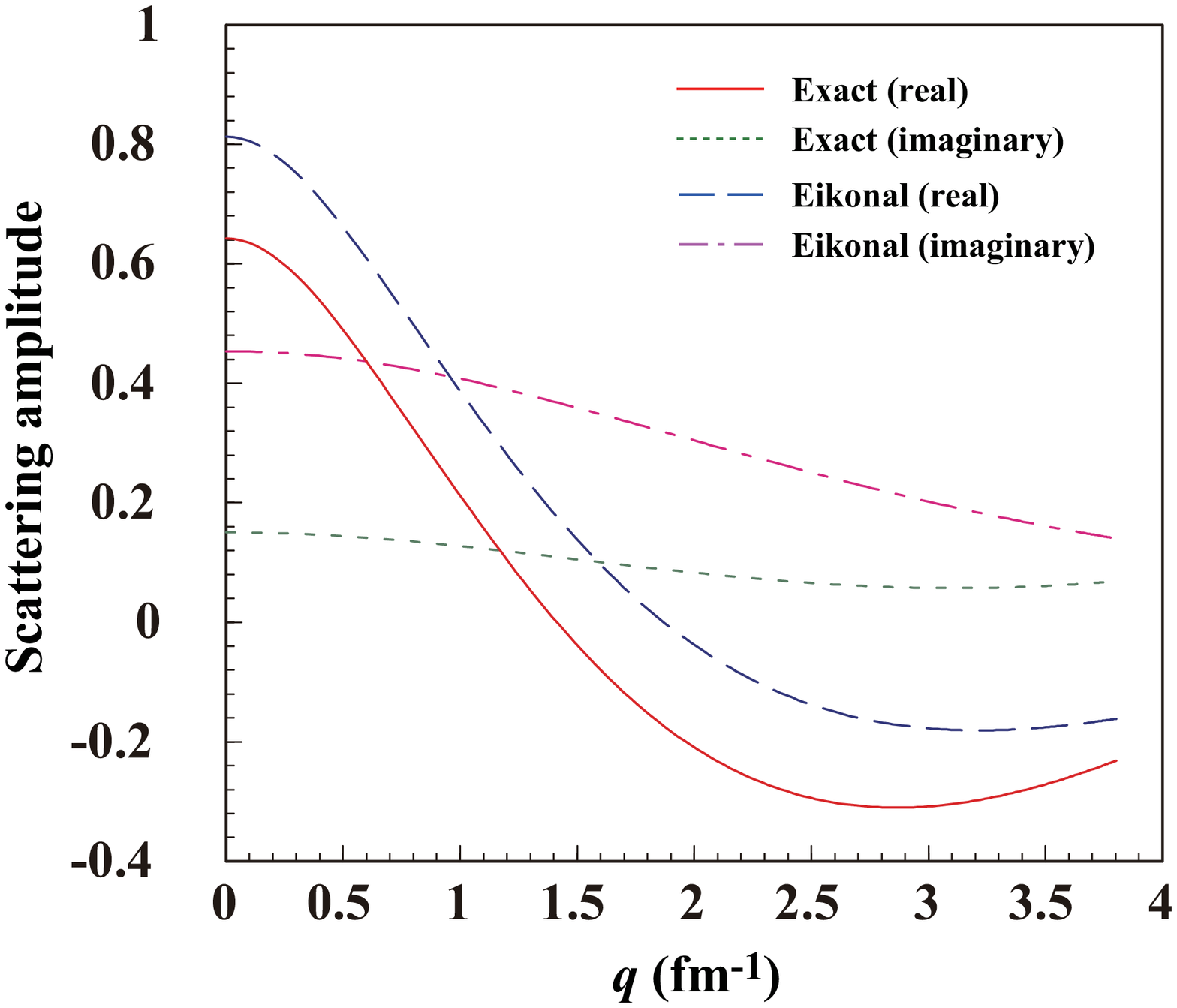}
\includegraphics[width=6.5 cm]{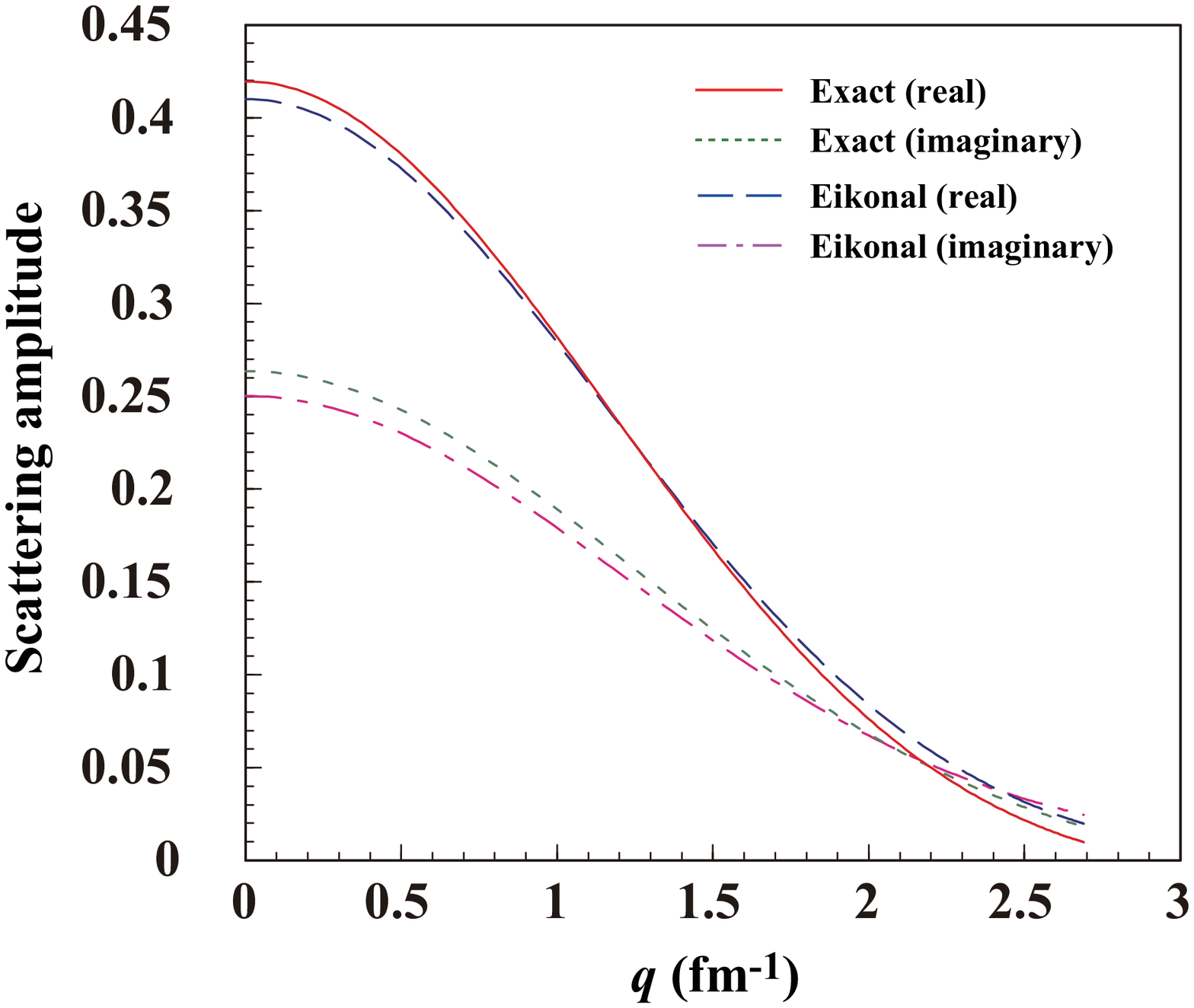}
\caption{
The on-shell NN scattering amplitude $f_{\rm {NN}}(\vq)$ 
calculated with the bare NN potential AV18 
at the laboratory energy $E_{\rm {NN}}=300$ MeV in the left panel and 
with the JLM $g$-matrix \cite{JLM} 
at the laboratory energy $E_{\rm {NN}}=150$ MeV in the right panel. 
The solid (dashed) and dotted (dash-dotted) lines show, respectively,
the real and imaginary parts of $f_{\rm NN}(\vq)$ of the exact (eikonal)
calculation. 
The left panel is taken from Ref.~\citen{Yahiro-Glauber}. 
}
\label{fig:f-NN}
\end{figure}

The Glauber model is based on the eikonal approximation 
for NN scattering and the eikonal and adiabatic approximations for 
nucleus-nucleus scattering. The condition for the eikonal approximation 
to be good is 
\bea
\vert v_{NN}(r_{NN})/E_{NN} \vert \ll 1, \quad  k_{NN} a \gg 1 \;,
\label{condition}
\eea
where $E_{NN}$  and $k_{NN}$ represent a kinetic energy 
and a wave number, respectively, 
of the relative motion between colliding two nucleons, 
and $a$ is a range of $v_{ij}$. 
This condition is not well satisfied for 
the realistic NN potential that has a strong short-ranged repulsive core 
at small relative distance ($r_{NN}$) between two nucleons; 
for example, $v_{ij} \sim 2000$ MeV at $r_{NN}=0$ for AV18~\cite{Wiringa}. 
Actually, the eikonal approximation is not good 
for NN scattering due to $v_{ij}$, as shown in the left panel of 
Fig.~\ref{fig:f-NN}. 
To avoid this problem, a slowly-varying function such as the Gaussian form 
is used in the profile function of the Glauber model~\cite{GM}. 
This procedure is justified in the following.

In general, the $g$-matrix has much milder $r$ dependence than 
the bare NN potential $v_{ij}$~\cite{Yahiro-Glauber}. For example, 
the JLM $g$-matrix \cite{JLM} keeps this property. This means that 
the $g$-matrix is more suitable than $v_{ij}$ as an input of 
the Glauber model. Actually, the eikonal approximation is quite good 
for NN scattering due to the JLM $g$-matrix, 
as shown in the right panel of Fig.~\ref{fig:f-NN}. 
The Glauber model~\cite{Glauber} is then more 
applicable to Eq.~\eqref{schrodinger-effective} than 
the original many-body Schr\"odinger equation with $v_{ij}$. 
In this case, the input of the Glauber model is not 
the profile function proposed in Ref.~\citen{GM} 
but the $g$-matrix~\cite{Yahiro-Glauber}. 
At higher incident energies where the Glauber model is used, 
the $g$-matrix is reduced to the $t$-matrix that has no medium effect. 
The $t$-matrix also has weak $r_{NN}$ dependence~\cite{Yahiro-Glauber}. 
This fact justifies the usage of the profile function 
with the Gaussian form.

%---figure 1 and 2-----------------------------------------------------
%----------------------------------------------------------------

\subsection{Application of double-folding model to reaction cross sections for 
Ne isotopes}
\label{Double-folding-model}

We analyze the scattering of Ne isotopes from a $^{12}$C target at 
240~MeV/nucleon. In the scattering the projectile breakup is weak, 
since the target is light and $E$ is large; see 
Sec.~\ref{Reorientation effect} for the point. 
The double-folding model becomes reliable in this situation. 
In the model, the potential $U$ between P and T  consists of 
the direct and exchange parts~\cite{DFM-standard-form,DFM-standard-form-2}. 
The exchange part is non-local, but it can be localized 
with the local semi-classical approximation~\cite{Brieva-Rook} in which
P is assumed to propagate as a plane wave with 
the local momentum within a short range of the 
NN interaction. The validity of this localization is shown 
in Ref.~\citen{Minomo:2009ds}.
As the $g$-matrix interaction we take the Melbourne interaction
\cite{Amos,von-Geramb} that is constructed from the Bonn-B 
NN potential \cite{BonnB}. 
The projectile densities are constructed by either (I) 
by antisymmetrized molecular dynamics (AMD)~\cite{Kimura} 
with the Gogny D1S interaction~\cite{Gogny,D1S} 
or by (II) the deformed Woods-Saxon (DWS) model~\cite{Sumi:2012fr} with 
the deformation evaluated by AMD. 
Model I has no adjustable 
parameter, but the density is inaccurate in the asymptotic region. 
Model II provides the density with the proper asymptotic form, 
but the model includes potential parameters. 
As the potential parameter set, we use the parameter set recently proposed 
by R.~Wyss~\cite{WyssPriv}.
This set is intended to reproduce spectroscopic properties of 
high-spin states from light to heavy deformed nuclei,
e.g., the quadrupole moment and the moment of inertia,
and at the same time the root mean square (RMS) radius 
crucial for the present analysis.

%%%%%%%%%%%%%%%%%%%%%%%
%%%  Figure
%%%%%%%%%%%%%%%%%%%%%%%
\begin{figure}[htbp]
\begin{center}
 \includegraphics[width=0.45\textwidth,clip]{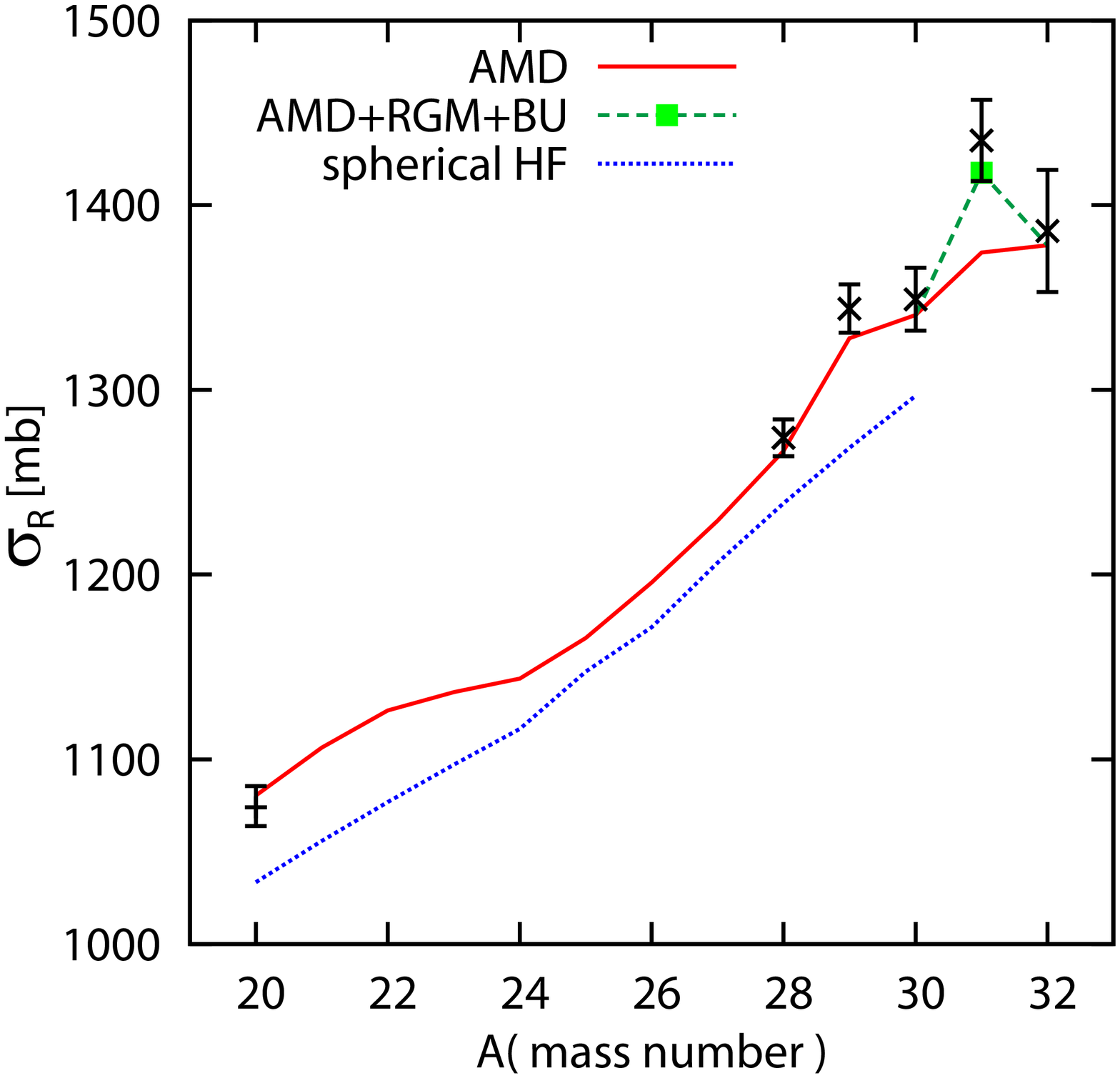}
 \includegraphics[width=0.45\textwidth,clip]{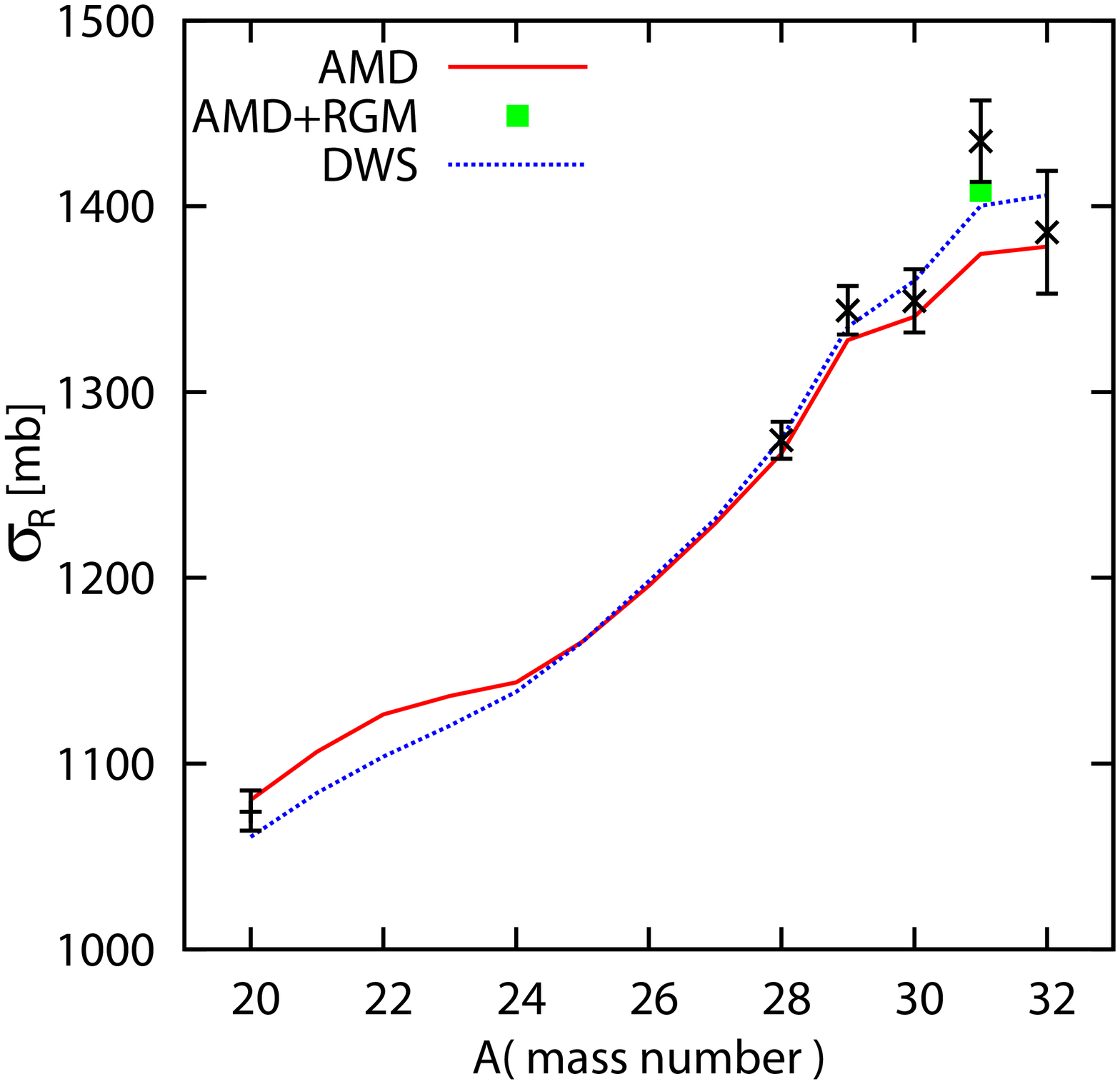}
 \caption{(Color online) Reaction cross sections for the scattering 
 of Ne isotopes on $^{12}$C at 240~MeV/nucleon. 
 The experimental data for $A=28-32$
 are taken from Ref.~\citen{Takechi}.
 The data for $^{20}$Ne is deduced
 from measured $\sigma_{\rm I}$ at around 1~GeV/nucleon~\cite{Ne20-sigmaI}
 with the Glauber model~\cite{Takechi}.
In the left panel, the solid (dotted) line represents the results 
of AMD (spherical Gogny-HF). A closed square 
is the result of AMD with the tail and breakup corrections.
In the right panel, the dotted line represents the results of the DWS model, 
while the solid line corresponds to the results of AMD. 
  }
 \label{Fig-AMD-HF-sigma-R}
\end{center}
\end{figure}

Figure~\ref{Fig-AMD-HF-sigma-R} represents the total reaction cross section 
$\sigma_{\rm R}$ for the scattering of Ne isotopes on a $^{12}$C target 
at 240~MeV/nucleon. As shown in the left panel, 
the AMD calculations (solid line) succeed in reproducing
the data~\cite{Takechi}, whereas 
the spherical Gogny Hartree-Fock (HF) calculation (dotted line) 
undershoots the data;
note that the spherical Gogny-HFB calculation yields the same result as
the spherical Gogny-HF calculation within the thickness of line.
The nuclei with $A > 30$ are unbound in these spherical calculations.
The enhancement from the dotted line to the solid line comes from
the deformation of the ground state, since the deformation is a main
difference between the two calculations.
The deformation increases the $\sigma_{\rm R}$ by at most 5\%.
The AMD results are consistent with all the data except $^{31}$Ne.
The underestimation of the AMD result for $^{31}$Ne comes from
the inaccuracy of the AMD density in its tail region.

The tail problem is solved by the following 
resonating group method (RGM)~\cite{Minomo:2011bb}.
In principle the ground state 
$\Phi(^{31}{\rm Ne}; 3/2^-_1)$ 
of $^{31}$Ne can be expanded in terms of
the ground and excited states $\Phi(^{30}{\rm Ne}; J^\pi_n)$ 
of $^{30}$Ne. This means that
the ground state of $^{31}$Ne is described by
the $^{30}$Ne+n cluster model with $^{30}$Ne excitations.
The cluster-model calculation can be done with RGM 
in which the ground and excited states of $^{30}$Ne are constructed 
by AMD. Here the wave function of $^{30}$Ne includes many excited
states with positive- and negative-parity below 10 MeV in excitation
energy. This AMD+RGM calculation is quite time consuming, 
but it is done for $^{31}$Ne. 
The tail correction to $\sigma_{\rm R}$ is 35~mb that corresponds 
to 2.5\% of $\sigma_{\rm R}$.
The reaction cross section with the tail correction 
(a square symbol) well reproduces the experimental 
data~\cite{Takechi} with no adjustable parameter. Thus 
$^{31}$Ne is a halo nucleus with large deformation. 
The DWS model~\cite{Sumi:2012fr} well simulates the result of 
AMD+RGM for $^{31}$Ne, as shown in the right panel 
of Fig.~\ref{Fig-AMD-HF-sigma-R}. 
This may suggest that the DWS model is a handy way of simulating 
the AMD calculation with the tail correction. 
The difference between the AMD model and 
the DWS model for $^{28-32}$Ne may show tail corrections to the AMD results.

\begin{figure}[htbp]
\begin{center}
 \includegraphics[width=0.5\textwidth,clip]{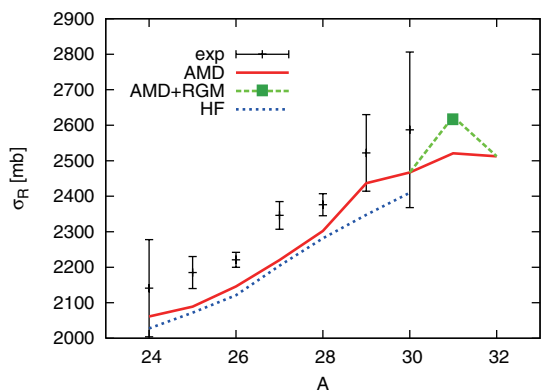}
 \caption{(Color online) Reaction cross sections for the scattering of Ne
isotopes on a $^{28}$Si target at 38-60~MeV/nucleon. 
 The experimental data are taken from Ref.~\citen{Khouaja}. 
 See Fig.~\ref{Fig-AMD-HF-sigma-R}  for the definition of lines. 
  }
 \label{Fig-AMD-HF-sigma-R-Low-E}
\end{center}
\end{figure}

The same analysis is made for the scattering of Ne 
isotopes on a $^{28}$Si target at 38-60~MeV/nucleon. 
As shown in Fig.~\ref{Fig-AMD-HF-sigma-R-Low-E}, the deformation effect is 
significant also for the lower incident energies, and consequently, 
the double-folding model yields better agreement 
with the experimental data~\cite{Khouaja}, 
where a normalization factor is multiplied in the theoretical results 
so as to reproduce measured $\sigma_{\rm R}$ of 
$^{12}$C+$^{12}$C scattering at 38-60~MeV/nucleon.

%%%%%%%%%%%%%%%%%%%%%%%
%%%  Figure
%%%%%%%%%%%%%%%%%%%%%%%

\subsection{Breakup, dynamical deformation and reorientation effects}
\label{Reorientation effect}

For a weakly bound system such as $^{31}$Ne, 
the projectile breakup effect is not perfectly negligible.
This effect is estimated by assuming the two-body model 
for the $^{30}$Ne+n system and solving the three-body dynamics 
of the $^{30}$Ne+n+$^{12}$C system with CDCC. 
Here the potential between $^{30}$Ne and $^{12}$C
and that between n and $^{12}$C are constructed with the double-folding model 
with the Melbourne $g$-matrix, 
and the potential between $^{30}$Ne and n is made
with the well-depth method; see Refs.~\citen{30Ne}
for the potential parameters.
The correction is 10~mb corresponding to 0.7\% of $\sigma_{\rm R}$.

When the projectile is deformed in the intrinsic frame, 
the deformation enlarges the radius of the projectile density 
in the space-fixed frame 
and eventually enhances the reaction cross section. 
This static deformation effect has already been included 
in the double-folding model by making the angular momentum projection. 
Another effect is the dynamical deformation effect, i.e., 
the effect of rotational motion of deformed projectile 
during the scattering. This effect on $\sigma_{\rm R}$ is small 
for intermediate-energy nucleus-nucleus scattering~\cite{Minomo-DWS}. 
This has been confirmed with the adiabatic 
approximation to the rotational motion of projectile and 
the eikonal approximation to the relative motion between projectile and 
target~\cite{Minomo-DWS}. 
In this subsection, the effect is estimated with no approximation. 
For this purpose, we consider 
the scattering of $^{30}$Ne from $^{12}$C at 240~MeV/nucleon and 
do coupled-channel calculations between the $0^{+}$ ground state 
and the  first $2^{+}$ state of $^{30}$Ne. 
The projectile density is calculated by the DWS model 
with the deformation evaluated by AMD. The coupling potentials 
in the coupled-channel calculations 
are obtained by the so-called single-folding model. Namely, 
the nucleon-$^{12}$C potential is first evaluated by folding 
the Melbourne-$g$-matrix interaction with the target density and 
the coupling potentials are obtained by folding 
the nucleon-$^{12}$C potential with the projectile transition densities.

In the single-channel calculation with no dynamical deformation effect, 
the resultant reaction cross section is 1469~mb. 
This result overestimates the corresponding result of the double-folding model 
by about $10~\%$, which is accurate enough for the present test. 
In the coupled-channel calculation, the resulting reaction cross section 
is 1468~mb. Thus the dynamical rotation effect on the reaction cross section 
is estimated as less than $0.1~\%$.  
The reason why the effect is small 
for intermediate-energy nucleus-nucleus scattering is shown in 
Ref.~\citen{Minomo-DWS}. 
The integrated inelastic cross section to the first $2^{+}$ state is 2.9~mb. 
This is 0.2~\% of $\sigma_{\rm R}$, indicating that 
$\sigma_{\rm I} \approx \sigma_{\rm R}$.

The folding potential $U$ is not spherical in general, when 
the spin of projectile is not zero. 
This reorientation effect is also tested by the coupled-channel
calculation for the scattering of $^{31}$Ne$(3/2^{-})$ from $^{12}$C at 
240~MeV/nucleon, where the single-folding model is used. 
The resultant reaction cross section is 1512 mb, whereas 
the corresponding cross section is 1515 mb 
when the non-spherical part of $U$ is switched off.
The reorientation effect is 0.2~\% and hence negligible
for intermediate-energy nucleus-nucleus scattering.  

\subsection{Summary on microscopic reaction theory}
\label{Summary-on-microscopic-reaction-theory}

We have constructed a microscopic reaction theory 
for nucleus-nucleus scattering, using the multiple scattering theory. 
This is an extension of the Kerman-McManus-Thaler theory 
for nucleon-nucleus scattering. In the theory, 
nucleus-nucleus scattering is described by a 
multiple scattering series due to the $g$-matrix NN interaction 
instead of the realistic one. The former has much milder 
$r$-dependence than the latter and therefore 
the Glauber model is applicable for this theory with high accuracy. 

In the nucleus-nucleus scattering from lighter targets at intermediate 
energies, the breakup effect is small and hence the double-folding model 
becomes reliable. In this situation, we can use the double-folding model 
in which projectile and target 
densities are constructed with fully microscopic structure theories such as 
AMD, HF and HFB. These fully microscopic theories have been applied 
to measured reaction cross sections for Ne isotopes. 
In the Island-of-inversion region, the nuclei are strongly deformed. 
In particular, $^{31}$Ne is a halo nucleus with strong deformation. 
The dynamical deformation effect and the reorientation effect are also 
found to be small, so that the interaction cross section is identical with 
the reaction cross section with high accuracy.

\section{Four-body CDCC}
\label{Four-body-CDCC}
 In this section we present a method of treating four-body scattering 
 in which the projectile breaks up into three constituents. 
 This method is called four-body CDCC. In this method, 
 the three-body continuum states of projectile are 
 discretized by diagonalizing the internal Hamiltonian
 of projectile with the Gaussian basis functions. This 
 discretization is called the pseudo-state method. 
 Recently, we proposed a novel method of calculating continuous breakup
 cross sections with the complex-scaling method~\cite{ABC,CSM1}. 
 This section is a brief review of Refs. 
 \citen{Matsumoto3,Matsumoto4,Matsumoto:2010mi}.   

\subsection{Formulation}

We consider the scattering of $^6$He as a typical example of four-body
breakup reactions. Here  we focus our discussion on the 
differential breakup cross section. 
The scattering of $^6$He on a target $A$ is described by 
the four-body Schr\"odinger equation 
\begin{align}
[H-E_{\rm tot}]|\Psi^{(+)}\rangle =0 
\label{eq:4b-Schr-ex}
\end{align}
with the outgoing boundary condition,
where the total energy $E_{\rm tot}$ satisfies 
$E_{\rm tot}=E_{\rm in}^{\rm CM}+\ve_0$  
for the corresponding incident energy $E_{\rm in}^{\rm CM}$ 
in the center-of-mass system and the ground-state energy $\ve_0$ of $^6$He.
The total Hamiltonian $H$ is defined by  
\begin{align}
 H &=K_R+h_{\rm P}+U_{nA}+U_{nA}+U_{\alpha A}+ V_{\rm \alpha A}^{\rm Coul}
\end{align}
with 
\begin{align}
 h_{\rm P}&=K_{y}+K_{r}+V_{nn}+V_{n\alpha}+V_{n\alpha},
\end{align}
where $h_{\rm P}$ is the internal Hamiltonian of $^6$He. 
The relative coordinate between $^6$He and A is denoted by 
$\bm{R}$ and  the internal coordinates of $^6$He are by 
a set of Jacobi coordinates, $\bm{\xi}=(\bm{y},\bm{r})$. 
Momenta conjugate to $\bm{R}$ and $(\bm{y},\bm{r})$ are
represented by $\bm{P}$ and  
$(\bm{p},\bm{k})$, respectively. The kinetic energy
operator associated with $\bm{R}$ ($\bm{\xi}$) is represented
by $K_{R}$ ($K_\xi$), $V_{xx'}$
is a nuclear plus Coulomb interaction between $x$ and $x'$, 
and $U_{xA}$ and 
$V_{xA}^{\rm Coul}$ are nuclear and Coulomb potentials
between $x$ and $A$, respectively.  

In CDCC with the pseudostate discretization method, 
the scattering is assumed to take place 
in a modelspace~\cite{Matsumoto, Egami,Matsumoto3, Matsumoto4}: 
\begin{eqnarray}
 {\cal P}=\sum_{\gamma}|\Phi_{\gamma}\rangle \langle \Phi_{\gamma}|, 
 \label{eq:com-set} 
\end{eqnarray}
where ${\Phi}_{\gamma}$ is the $\gamma$-th eigenstate obtained by 
diagonalizing $h_{\rm P}$ with $L^2$-type basis functions. 
The four-body Schr\"odinger equation is then solved 
in the modelspace: 
\begin{align}
{\cal P}[H-E_{\rm tot}]{\cal P}|\Psi^{(+)}_{\rm CDCC} \rangle =0.
\label{eq:4b-Schr}
\end{align}
The modelspace assumption has already been justified by the fact that 
calculated elastic and breakup cross sections converge with respect to
extending 
the modelspace~\cite{Matsumoto, Egami, Matsumoto3, Matsumoto4}. 

The exact $T$-matrix element to a breakup state with $(\bm{p},\bm{k})$ 
can be described by 
\begin{eqnarray}
T_{\ve}(\bm{p},\bm{k},{\bm{P}})=
 \langle \psi^{(-)}_\ve(\bm{p},\bm{k}) \chi^{(-)}_\ve(\bm{P})|
U-V^{\rm Coul}_{\rm ^6He}
|\Psi^{(+)}\rangle
, \label{Tmat0}
 \label{exact-T}
\end{eqnarray}
where 
$V_{\rm ^6He}^{\rm Coul}$ is the Coulomb interaction between $^6$He
and $A$. The final-state wave functions,
$|\psi_\ve^{(-)}(\bm{p},\bm{k})\rangle$ and   
$|\chi_\ve^{(-)}(\bm{P})\rangle$, with the incoming boundary condition, 
are defined by  
\begin{eqnarray}
 \left[T_R+V^{\rm Coul}_{\rm ^6He}-(E_{\rm tot}-\ve)\right]
  |\chi^{(-)}_\ve(\bm{P})\rangle &=&0,\\
 \left[h_{\rm P}-\ve\right]|\psi_\ve^{(-)}(\bm{p},\bm{k}) \rangle&=&0, 
  \label{eq:Sch-B}
\end{eqnarray}
where 
$E_{\rm tot}-\ve=(\hbar P)^2/(2\mu_R)$
and
$\ve=(\hbar p)^2/(2\mu_{y})+(\hbar k)^2/(2\mu_{r})$ 
for reduced masses $\mu_{R}$ and $\mu_{\xi}$ of coordinates $\bm{R}$ and
$\bm{\xi}$, respectively. 
Inserting the approximate complete set Eq. \eqref{eq:com-set} 
into Eq. \eqref{exact-T}, one can get the $T$-matrix element 
with high accuracy~\cite{Matsumoto, Egami, Matsumoto3, Matsumoto4}, 
\begin{eqnarray}
T_{\ve}(\bm{p},\bm{k},{\bm{P}}) 
\approx \sum_{\gamma \ne 0}
\langle \psi_\ve^{(-)}(\bm{p},\bm{k}) |\Phi_{\gamma}\rangle
T_{\gamma}
\label{approx-T}
\end{eqnarray}
with the CDCC $T$-matrix element
\begin{eqnarray}
T_{\gamma}=\langle \Phi_{\gamma}\chi^{(-)}_{\ve_\gamma}(\bm{P}_\gamma)|
 U-V^{\rm Coul}_{\rm ^6He}
 |\Psi^{(+)}_{\rm CDCC}\rangle  
\end{eqnarray}
to the $\gamma$-th discrete breakup state $\Phi_{\gamma}$ with eigenenergy
$\ve_\gamma$.
Here Eq.~\eqref{approx-T} is derived by replacing $\bm{P}$ by $\bm{P}_\gamma$
in $\chi^{(-)}_\ve(\bm{P})$.
The $T_{\gamma}$ are obtainable with CDCC, but 
it is quite hard to calculate 
the smoothing factor 
$\langle\psi_\ve^{(-)}(\bm{p},\bm{k})|\Phi_{\gamma}\rangle $ directly with 
either numerical integration~\cite{LSCSM1} or the complex-scaling 
method~\cite{LSCSM2}. 
Hence, we propose a new way of obtaining 
the differential cross section with respect to $\ve$ without 
calculating the smoothing factor. 

Using Eq. \eqref{approx-T}, one can rewrite 
the differential cross section into 
\begin{eqnarray}
\frac{d^2\sigma}{d\ve d\Omega_{\bm{P}}} &=& 
\int d \vp' d \vk' 
\delta(\ve-\ve')
|T_{\ve'}(\vp',\vk',{\bm{P}'})|^2
\approx \frac{1}{\pi}{\cal R}(\ve,\Omega_{\bm{P}})
\label{xsec-1}
\end{eqnarray}
with the generalized response function 
\begin{eqnarray}
 {\cal R}(\ve,\Omega_{\bm{P}})
={\rm Im} \Big[
\sum_{\gamma,\gamma'\ne 0} T_{\gamma}^{*} 
\langle \Phi_{\gamma}|G^{(-)}|\Phi_{\gamma'}\rangle
T_{\gamma'} 
\Big] 
,\label{response-fun}
\end{eqnarray}
where 
 $\displaystyle G^{(-)}=\lim_{\eta \to +0} (\ve-h_{\rm P} - i\eta)^{-1}$.
There is no smoothing factor in Eq. \eqref{response-fun}, as expected. 
The propagator $G^{(-)}$ operates only 
on spatially damping functions $\Phi_{\gamma}$. This makes the 
calculation of $\langle \Phi_{\gamma}|G^{(-)}|\Phi_{\gamma'}\rangle$ feasible. 

In order to calculate $\langle \Phi_{\gamma}|G^{(-)}|\Phi_{\gamma'}\rangle$, 
we use the complex-scaling method in which 
the scaling transformation operator $C(\theta)$ and its inverse are defined by 
\begin{eqnarray}
\langle \vrr,\vy | C(\theta)|f \rangle 
&=& e^{3i\theta}f(\vrr e^{i\theta},\vy e^{i\theta}), \\
\langle f | C^{-1}(\theta)| \vrr,\vy \rangle 
 &=& \{ e^{-3i\theta}f(\vrr e^{-i\theta},\vy e^{-i\theta}) \}^{*}.
\end{eqnarray}
Using the operators, one can get 
\begin{eqnarray} 
\langle \Phi_{\gamma}|G^{(-)}|\Phi_{\gamma'}\rangle
=\langle \Phi_{\gamma}|C^{-1}(\theta) G_{\theta}^{(-)}C(\theta)|
\Phi_{\gamma'}\rangle ,
\label{C-G-1}
\end{eqnarray}
where 
\begin{eqnarray}
G_{\theta}^{(-)}=\lim_{\eta \to +0} \frac{1}{\ve-h^{\theta}_{\rm P} - i\eta} . 
\end{eqnarray}
with 
$h_{\rm P}^\theta=C(\theta)h_{\rm P}C^{-1}(\theta)$.
When $-\pi<\theta<0$,
the scaled propagator $\langle\bm{\xi}|G^{(-)}_\theta|\bm{\xi}'\rangle$ 
is a damping function of $\bm{\xi}$ and $\bm{\xi}'$; 
note that $\theta$ is negative 
since $G^{(-)}$ has the incoming boundary condition.
The scaled propagator can be expanded with $L^2$-type basis functions 
with high accuracy:
\begin{eqnarray}
G^{(-)}_\theta &\approx& \sum_i
  \frac{|\phi^\theta_i\rangle\langle\tilde{\phi}_i^\theta|}
  {\ve-\ve_i^\theta},
  \label{Gtheta2}
\end{eqnarray}
where $\phi_i^\theta$ is the $i$-th eigenstate of 
$h_{\rm P}^\theta$ in a modelspace spanned by $L^2$-type basis functions,
$\langle\tilde{\phi}_i^\theta|h_{\rm P}^\theta
|\phi_{i'}^\theta\rangle=
\ve_i^\theta \delta_{ii'}$. 
Inserting Eq. \eqref{Gtheta2} into Eq. \eqref{response-fun} through 
Eq. \eqref{C-G-1} 
leads to a useful form of 
\begin{eqnarray}
 \frac{d^2\sigma}{d\ve d\Omega_{\bm{P}}}&\approx&
  \frac{1}{\pi}{\rm Im}\sum_i\frac{T_{i}^\theta \tilde{T}_i^\theta}
  {\ve-\ve_i^\theta}
  \label{sigma-CSM}
\end{eqnarray}
with 
\begin{eqnarray}
 \tilde{T}_i^\theta \equiv
 \sum_{\gamma'} \langle \tilde{\phi}_i^\theta|C(\theta)
 |\Phi_{\gamma'} \rangle 
 T_{\gamma'}, &&\;\;
% \label{R-7} \\
  T_i^\theta\equiv
  \sum_\gamma
  T_\gamma^*   \langle \Phi_\gamma|C^{-1}(\theta)|\phi_i^\theta \rangle.
  \label{R-8}
\end{eqnarray}
This method does not require to calculate 
the exact three-body continuum states $\psi_{\ve}^{(-)}(\vk,\vp)$. The
convergence of this method is shown in Ref.~\citen{Matsumoto:2010mi}.

\subsection{Differential breakup cross section for $^6$He scattering}

%%%%%%%%%%%%%%%%%%%%%%%
%%%  Figure 5
%%%%%%%%%%%%%%%%%%%%%%%
\begin{figure}[htbp]
\begin{center}
 \includegraphics[width=0.45\textwidth,clip]{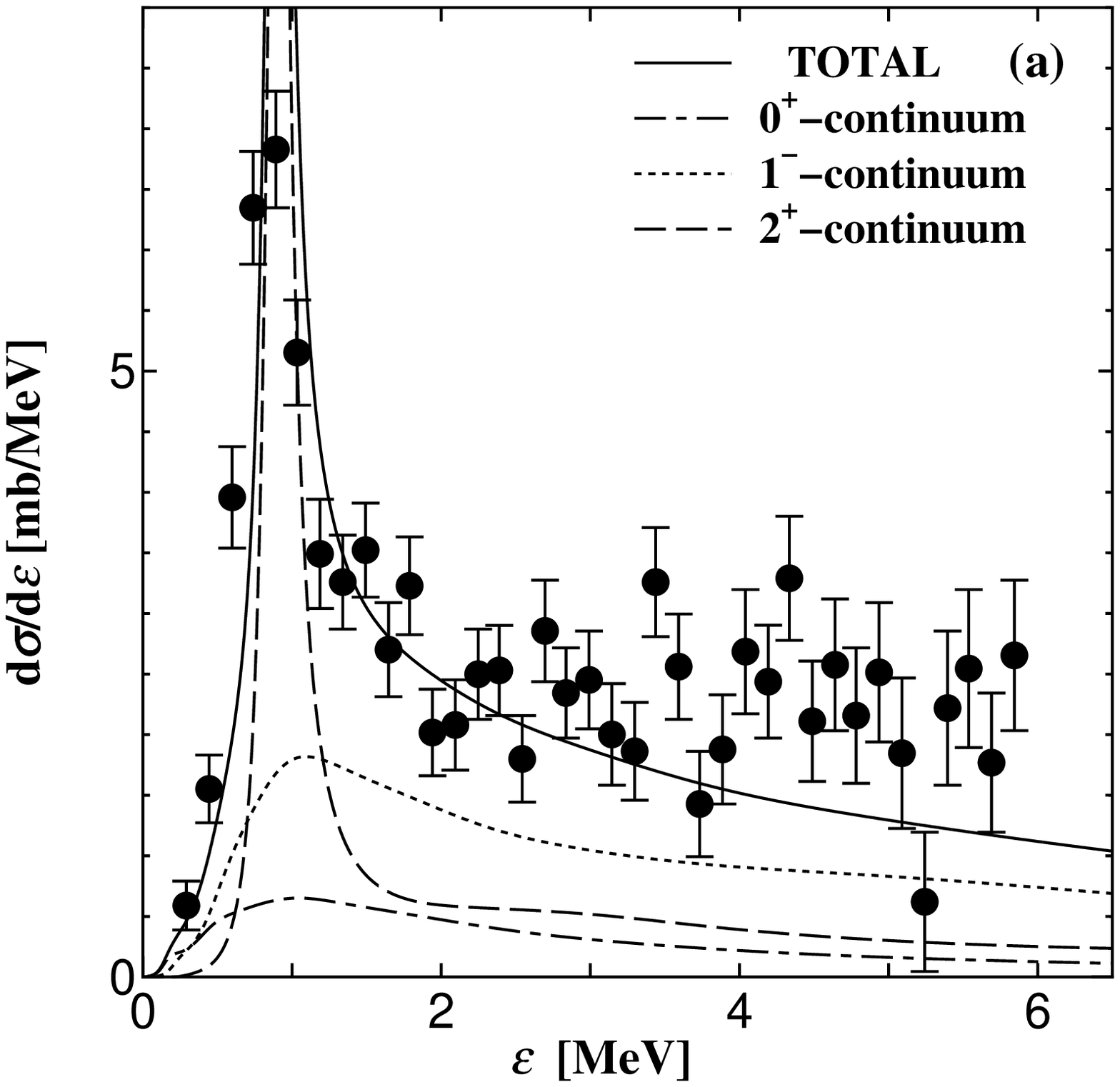}
 \includegraphics[width=0.45\textwidth,clip]{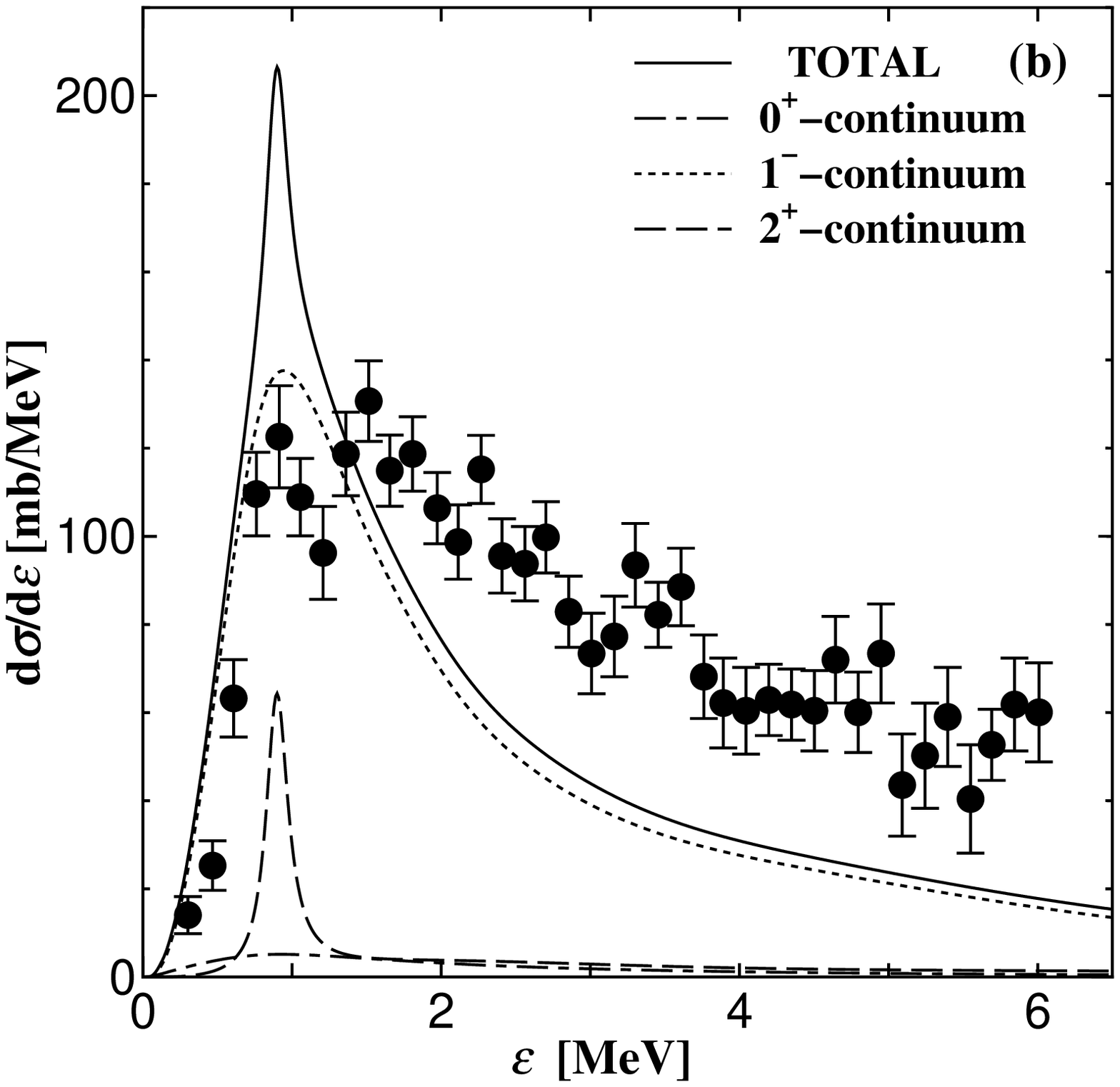}
 \caption{Breakup cross sections for (a) $^6$He+$^{12}$C 
 scattering at 240 MeV/A and (b) $^6$He+$^{208}$Pb scattering at 240 MeV/A. 
 The solid lines are results of full-fledged the four-body CDCC calculations. 
 The dot-dashed, dotted, and dashed lines correspond to contributions of $0^+$,
 $1^-$, and $2^+$ breakup, respectively. 
 The experimental data are taken
 from Ref.~\citen{Aumann}.} 
 \label{results}
\end{center}
\end{figure}

In Fig.~\ref{results}, the calculated breakup cross sections are
compared with the experimental data~\cite{Aumann} on  
$^6$He+$^{12}$C and $^6$He+$^{208}$Pb reactions at 240 MeV/A. 
In the calculation, we take the same potentials as in Ref.~\citen{4BEIK}
for $n$-$^{208}$Pb and $\alpha$-$^{208}$Pb subsystems. The optical
potential for $n$-$^{12}$C subsystem is taken from the global
nucleon-nucleus potential~\cite{Koning}, while the optical potential for  
$\alpha$-$^{12}$C subsystem is constructed from the $^{12}$C+$^{12}$C 
potential at 200 MeV/A~\cite{12C-12C} by changing the radius parameter 
from $^{12}$C to $\alpha$. Nuclear breakup is dominant for
$^6$He+$^{12}$C scattering at 240 MeV/A, while Coulomb breakup 
to $1^-$ continuum is dominant for $^6$He+$^{208}$Pb scattering. 
For $^{12}$C target, the present theoretical result is consistent 
with the experimental data except for the peak of the 2$^+$-resonance
around $\ve=1$ MeV. Similar overestimations are also seen in 
the results of four-body distorted-wave Born approximation (DWBA)~\cite{4BDW}. 
For $^{208}$Pb target, the present method 
underestimates the experimental data at $\ve \ga 2$~MeV. A possible
origin of this underestimation is that the inelastic breakup reactions
are not included in the present calculation.  
It has been reported in Ref.~\citen{4BDW} 
that the inelastic breakup effect is not negligible and thereby 
the elastic breakup cross section calculated with 
four-body DWBA underestimates the data. 

\subsection{Summary on four-body CDCC}

We have proposed a new version of CDCC for treating four-body breakup. 
This method is called four-body CDCC. In the method, 
the three-body continuum of projectile is treated with 
the complex-scaling method. 
The validity of this method is checked for not only 
the elastic scattering but also the breakup reactions of $^6$He. 
Clear convergence with respect to expanding the modelspace is seen 
in both the elastic and the breakup cross section. 
Four-body CDCC with the complex-scaling method is indispensable 
to study properties of unstable nuclei with two-neutron halo structure.

\section{Eikonal reaction theory}
\label{Eikonal-reaction-theory}

 In this section, we present an accurate method of treating
neutron removal reactions at intermediate incident energies 
as an extension of CDCC and the Glauber model. 
This method is referred to as the eikonal reaction theory. 
This section is a brief review of Refs.~\citen{ERT,Hashimoto:2011nc}.

\subsection{Formulation}
  Let us assume that a projectile (P) consists 
 of a core nucleus (c)  and a neutron (n). 
 The scattering of P on a target (T) is then described by 
the three-body (c+n+T) Schr\"odinger equation
\beq
   \left[ -\frac{\hbar^2}{2\mu}\nabla_R^2+ h + U(r_{\rm c},r_{\rm n})-E \right]\Psi=0
\label{Schrodinger-eq}
\eeq
with the interaction
\beq
 U = U_{\rm n}^{\rm (N)}(r_{\rm n})
 + U_{\rm c}^{\rm (N)}(r_{\rm c}) + U_{\rm c}^{\rm (C)}(r_{\rm c}) ,
\label{Pot}
\eeq
where $h=T_r +V(\vrr)$ is the projectile Hamiltonian, 
$\vR=(\vb,Z)$ stands for the coordinate between P and T, 
$\vrr_{\rm x}$ (x=n or c) represents the coordinate between x and A, and 
$U_{\rm x}^{\rm (N)}$ and $U_{\rm c}^{\rm (C)}$ are 
the nuclear and Coulomb parts of the optical potential between x and T, 
respectively. Solving \eqref{Schrodinger-eq} with the eikonal approximation, 
one can get the $S$-matrix operator 
\bea
S=\exp\Big[
-i{\cal P}\int_{-\infty}^{\infty} dZ {\hat O^{\dagger}}U{\hat O}
\Big]  
\label{S-matrix-operator}
\eea
for the operator
\bea
{\hat O} =\frac{1}{\sqrt{\hbar {\hat v}}} e^{i {\hat K} \cdot Z}
\eea
with the wave-number operator ${\hat K}=\sqrt{2\mu(E-h)}/{\hbar}$
and the velocity operator ${\hat v}={\hbar {\hat K}}/{\mu}$
of the relative motion between P and T, where 
${\cal P}$ is the path ordering operator. 
In the Glauber model, the adiabatic approximation is made as the secondary
approximation in which $h$ is replaced by the ground-state 
energy $\epsilon_0$, and hence ${\hat O^{\dagger}}U{\hat O}$ and
${\cal P}$ in \eqref{S-matrix-operator} are reduced to
$U/(\hbar v_0)$ and 1, respectively, where $v_0$ is the velocity of P
in the ground state relative to T.

The operator ${\hat O}$ shows internal motions of c and n 
during the scattering. 
The movement is small for short-range nuclear interaction, but not for 
long-range Coulomb interaction. 
This means that $U_{\rm n}^{\rm (N)}$ is commutable with
${\hat O}$ with high accuracy: 
\beq
    {\hat O^{\dagger}}U_{\rm n}^{\rm (N)} {\hat O} \rightarrow
     U_{\rm n}^{\rm (N)}/(\hbar v_0) .
\label{replacement-N}
\eeq
For the scattering of $^{31}$Ne($1p3/2$) from a $^{208}$Pb target at 
240~MeV/nucleon, 
the error due to the replacement \eqref{replacement-N} 
is estimated with CDCC; it is 0.2\% for 
the reaction cross section $\sigma_{\rm R}$, 1.9\% for 
the breakup cross section $\sigma_{\rm bu}$, 
4.1\% for the stripping cross section $\sigma_{\rm str}$. 
Using this replacement, one can get an important result
\beq
  S = S_{\rm n}S_{\rm c}
  \label{S-separation}
\eeq
with 
\bea
  S_{\rm n}&=&
    \exp\Big[   - \frac {i}{\hbar v_0} \int_{-\infty}^{\infty} dZ 
    U_{\rm n}^{\rm (N)} \Big] ,
  \label{Sn} \\
  S_{\rm c}&=&\exp\Big[
-i{\cal P}\int_{-\infty}^{\infty} dZ {\hat O^{\dagger}}(U_{\rm c}^{\rm (N)}+U_{\rm c}^{\rm (C)})
{\hat O} \Big] .
\label{Sc}
\eea
Thus $S$ can be separated into the neutron part $S_{\rm n}$ and
the core part $S_{\rm c}$.
One can not calculate $S_{\rm c}$ directly with Eq.~\eqref{Sc},
because it includes the operators ${\hat O}$ and ${\cal P}$.
However, $S_{\rm c}$ is the solution to the Schr\"odinger equation
\beq
   \left[ -\frac{\hbar^2}{2\mu}\nabla_R^2 + h + U_{\rm c}^{\rm (N)}(r_{\rm c})
   + U_{\rm c}^{\rm (C)}(r_{\rm c})-E \right]\Psi_{\rm c}=0, 
\label{Schrodinger-eq-core}
\eeq
when the eikonal approximation is made. One can obtain 
$S_{\rm c}$ by solving Eq.~\eqref{Schrodinger-eq-core} 
with eikonal-CDCC~\cite{Eikonal-CDCC} in 
which the eikonal approximation is made in the framework of CDCC. 
Non-eikonal corrections to $S_{\rm c}$ can be easily made by solving 
Eq.~\eqref{Schrodinger-eq-core} with CDCC instead of eikonal-CDCC, 
although it is not necessary for the present intermediate scattering. 
As mentioned above, $S_{\rm n}$ is obtained from Eq.~\eqref{Sn}. 
We can derive several kinds of cross sections with the product form
\eqref{S-separation}, following the formulation on the cross sections in
the Glauber model~\cite{Hussein,Hencken}.

\subsection{One-neutron removal cross section for $^{31}$Ne scattering}

The eikonal reaction theory is applied to one-neutron removal reactions 
for $^{31}$Ne+$^{12}$C scattering at 230~MeV/nucleon and
$^{31}$Ne+$^{208}$Pb scattering at 234~MeV/nucleon.
Table \ref{table1} presents several kinds of cross sections and
the spectroscopic factor 
${\cal S}=\sigma^{\rm exp}_{\rm -n}/\sigma^{\rm th}_{\rm -n}$. 
Thus ${\cal S}[1p3/2]$ little depends on the target and less than 1,
but ${\cal S}[0f7/2]$ does not satisfy these conditions.
In Ref.~\citen{Nakamura}, the Coulomb component
of $\sigma_{\rm -n}[^{208}{\rm Pb}]$ for a $^{208}$Pb target
is estimated to be 540~mb from the experimental values of
$\sigma_{\rm -n}[^{208}{\rm Pb}]$ and $\sigma_{\rm -n}[^{12}{\rm C}]$.
In the eikonal reaction theory, the Coulomb component
of $\sigma_{\rm -n}[^{208}{\rm Pb}]$ agrees with
$\sigma_{\rm bu}[^{208}{\rm Pb}]$ with good accuracy.
The spectroscopic factor evaluated from the Coulomb component is
${\cal S'}=540/\sigma^{\rm th}_{\rm bu}=0.675$ for the $1p3/2$ orbit and
7.36 for the $0f7/2$ orbit. Thus ${\cal S'}$ is consistent
with ${\cal S}$ only for the $1p3/2$ orbit.
Hence, we can infer that 
the major component of the $^{31}{\rm Ne}_{\rm g.s.}$ wave function
is $^{30}{\rm Ne}(0^+) \otimes 1p3/2$ (${\cal S} \sim 0.69$).
We adopt in the following this configuration.

\begin{table}
\begin{center}
\caption
{Several kinds of cross sections and the spectroscopic factors for
$^{31}$Ne+$^{12}$C scattering at 230~MeV/nucleon and
$^{31}$Ne+$^{208}$Pb scattering at 234~MeV/nucleon.
The cross sections are presented in units of mb and the data are taken from
Ref.~\citen{Nakamura}.
}
\label{table1}
\begin{tabular}{cccccccc}
\hline \hline
 & \multicolumn{3}{c}{$^{12}$C target} & \hspace{2mm} &
 \multicolumn{3}{c}{$^{208}$Pb target} \\ \cline{2-4}  \cline{6-8}
 {} & ~~~$p_{3/2}$~~~ & ~~~$f_{7/2}$~~~ & ~~~Exp.~~~ &
 {} & ~~~$p_{3/2}$~~~ & ~~~$f_{7/2}$~~~ & ~~~Exp.~~~ \\ \hline
 $\sigma_{\rm R}$ & 1572.5 & 1489.9 & & & 5518.0 & 4589.5 & \\
 $\sigma_{\rm bu}$ & 23.3 & 3.3 & & & 799.5 & 73.0 & (540) \\
 $\sigma_{\rm R}$(-n) & 1463.5 & 1458.6 & & & 5151.5 & 4524.2 & \\
 $\sigma_{\rm bu}$(-n) & 4.5 & 1.0 & & & 677.2 & 60.5 & \\ \hline
 $\sigma_{\rm str}$ & 90 & 29 & & & 244 & 53 & \\
 $\sigma_{\rm -n}$ & 114 & 32 & 79 & & 1044 & 126 & 712 \\
 ${\cal S}$ & 0.693 & 2.47 & & & 0.682 & 5.65 & \\ \hline \hline
\end{tabular}
\end{center}
\end{table}

The potential $V$ between c and n is not well known. Hence, ${\cal S}$ has
a theoretical error coming from the potential ambiguity. The error
is often estimated by changing the potential parameters by 30\%.
When the one-neutron separation energy $B_{\rm n}$ of $^{31}$Ne is 0.33~MeV,
${\cal S}= 0.693 \pm 0.133 \pm 0.061$
for $^{12}$C target and $0.682 \pm 0.133 \pm 0.062$
for $^{208}$Pb target, where the second and third numbers
following the mean value
stand for the theoretical and experimental uncertainties, respectively.
Thus ${\cal S}$ includes a sizable theoretical error.
This situation completely changes if we look at the
asymptotic normalization coefficient (ANC) $C_{\rm ANC}$~\cite{Xu}. 
When $B_{\rm n}=0.33$~MeV,
$C_{\rm ANC}= 0.320 \pm 0.010 \pm 0.028$~fm$^{-1/2}$
for $^{12}$C target and
$0.318 \pm 0.008 \pm 0.029$~fm$^{-1/2}$
for $^{208}$Pb target.
Thus, $C_{\rm ANC}$  has much smaller theoretical errors than
${\cal S}$. This means that the one-nucleon removal reaction is quite
peripheral.

\subsection{The eikonal reaction thoery for two-neutron removal}

The eikonal reaction theory is applicable for two-neutron removal reactions. 
Here we show the application for the scattering of $^6$He from 
$^{12}$C and $^{208}$Pb targets at 240 MeV/nucleon. 
In this case, the projectile is a three-body system and hence 
four-body CDCC should be used. 
The potentials for n-target and $\alpha$-target subsystems  
are calculated by the folding procedure shown 
in Sec.~\ref{Double-folding-model}. 
The wave functions are obtained by spherical Gogny-HF calculations. 
Table \ref{table2} shows the integrated cross sections 
for two-neutron removal of $^6$He. 
Our results are almost consistent with the experimental data~\cite{Aumann}. 
Thus we can clearly see the reliability of the eikonal reaction theory 
for two-neutron removal reactions on both light and heavy targets. 
\begin{table}[htb]
\begin{center}
\caption{Integrated cross sections for two-neutron removal of $^6$He. 
The cross sections are presented in units of mb and 
the experimental data are taken from Ref.~\citen{Aumann}. }
\label{table2}
\begin{tabular}{cccccc}
\hline \hline
 & \multicolumn{2}{c}{$^{12}$C target} & \hspace{2mm} &
 \multicolumn{2}{c}{$^{208}$Pb target} 
\\ \cline{2-3} \cline{5-6}
 {} & Calc. & Exp. &
 {} & Calc. & Exp. 
\\ \hline
% $\sigma_{\rm R}^{}$ & 652.6 & & & 3503.1 & \\
% $\sigma_{\rm bu}^{}$ & 16.1 & (30 $\pm$ 5) & & 514.1 & (650 $\pm$ 110) \\ 
 $\sigma_{1n~\rm str}^{}$ & 153.4 & 127 $\pm$ 14 & & 353.6 & 320 $\pm$ 90 \\
 $\sigma_{2n~\rm str}^{}$ & 29.0 & 33 $\pm$ 23 & & 148.9 & 180 $\pm$ 100 \\ 
 $\sigma_{\rm -2n}^{}$ & 198.5 & 190 $\pm$ 18 & & 1016.6 & 1150 $\pm$ 90 \\ 
\hline \hline
\end{tabular}
\end{center}
\end{table}

\subsection{Summary on the eikonal reaction theory} 
We have presented an accurate method of treating
neutron removal reactions at intermediate energies. 
In the theory, the nuclear and Coulomb breakup processes are accurately and
consistently treated by CDCC without making the adiabatic approximation 
to the latter, so that the removal cross section calculated with the method 
never diverges even in the presence of the Coulomb interaction.
For lower incident energies where 
the eikonal approximation is not perfectly accurate, 
one should make non-eikonal corrections to inclusive cross sections. 
This can be done easily by using CDCC instead of eikonal-CDCC.

$C_{\rm ANC}$ and ${\cal S}$ of the last neutron in $^{31}$Ne
are evaluated from the measured one-neutron removal reaction. 
$C_{\rm ANC}$ has a smaller theoretical error and 
weaker target-dependence than ${\cal S}$. Thus, $C_{\rm ANC}$ is determined
more accurately than ${\cal S}$. 
When the last neutron of $^{31}$Ne is in the $1p3/2$ orbit,
${\cal S} < 1$ for $B_{\rm n} \la 0.6$~MeV, and ${\cal S}$ and $C_{\rm ANC}$
have weaker target dependence.
When the last neutron is in the $1f7/2$ orbit, meanwhile,
${\cal S} > 1$ and ${\cal S}$ and $C_{\rm ANC}$
have stronger target dependence. 
These results indicate that the last neutron is mainly in the $1p3/2$ orbit. 
This means that $^{31}$Ne is deformed. This is consistent 
with our result in Sec.~\ref{Double-folding-model}. 
The accuracy of the Glauber model is systematically investigated for 
deuteron scattering at 200~MeV/nucleon; 
see Ref.~\citen{Hashimoto:2011nc} for the details.

\section*{Acknowledgements}

This review is based on the collaboration with 
K. Ogata, M. Kimura, Y. R. Shimizu, S. Hashimoto, 
M. Kawai, K. Kat\=o, M. Kohno and S. Chiba. 
The authors deeply appreciate their collaboration. 
The authors thank D. Baye, P. Descouvemont, K. Hagino, Y. Suzuki 
and Y. Sakuragi for useful discussions.

%\appendix
%\section{First Appendix} %Empty argument \section{} yields `Appendix'. 
%
%\section{Second Appendix}

%%--------------------------------------------------------------------%%
%%                           References                               %%
%%--------------------------------------------------------------------%%

%\bibliographystyle{h-physrev}
%\bibliography{bib-TM1}

\begin{thebibliography}{10}

\bibitem{Glauber}
R.~Glauber,
\newblock {\it in Lectures in Theoretical Physics} (Interscience, New York,
  1959) {\bf {\rm vol}. 1}, 315 (1959).

\bibitem{Yahiro-Glauber}
M.~Yahiro, K.~Minomo, K.~Ogata, and M.~Kawai,
\newblock Prog. Theor. Phys. {\bf 120}, 767 (2008).

\bibitem{Hussein}
M.~S. Hussein and K.~W. McVoy,
\newblock Nucl. Phys. A {\bf 445}, 124 (1985).

\bibitem{Hencken}
K.~Hencken, G.~Bertsch, and H.~Esbensen,
\newblock Phys. Rev. C {\bf 54}, 3043 (1996).

\bibitem{Gade}
A.~Gade {\em et~al.},
\newblock Phys. Rev. C {\bf 77}, 044306 (2008).

\bibitem{Ogawa01}
K.~Yabana, Y.~Ogawa, and Y.~Suzuki,
\newblock Nucl. Phys. A {\bf 539}, 295 (1992); Y. Ogawa, K. Yabana, and Y.
  Suzuki, Nucl. Phys. A {\bf 543}, 722 (1992); Y. Ogawa, T. Kido, K. Yabana,
  and Y. Suzuki, Prog. Theor. Phys. Suppl. {\bf 142}, 157 (2001).

\bibitem{Tostevin}
J.~Al-Khalili and J.~Tostevin,
\newblock Phys. Rev. Lett. {\bf 76}, 3903, (1996); J.A. Tostevin and B.A.
  Brown, Phys. Rev. C {\bf 74}, 064604 (2006).

\bibitem{Bertulani-92}
C.~A. Bertulani and K.~W. McVoy,
\newblock Phys. Rev. C {\bf 46}, 2638 (1992).

\bibitem{Bertulani-04}
C.~A. Bertulani and P.~G. Hansen,
\newblock Phys. Rev. C {\bf 70}, 157 (2001).

\bibitem{Ibrahim}
B.~Abu-Ibrahim and Y.~Suzuki,
\newblock Prog. Theor. Phys. {\bf 112} 1013, (2004); B. Abu-Ibrahim and Y.
  Suzuki, Prog. Theor. Phys. {\bf 114}, 901 (2005).

\bibitem{Capel-08}
P.~Capel, D.~Baye, and Y.~Suzuki,
\newblock Phys. Rev. C {\bf 78}, 054602 (2008).

\bibitem{CDCC-review1}
M.~Kamimura {\em et~al.},
\newblock Prog. Theor. Phys. Suppl. {\bf 89}, 1 (1986).

\bibitem{CDCC-review2}
N.~Austern {\em et~al.},
\newblock Phys. Rep. {\bf 154}, 125 (1987).

\bibitem{CDCC-foundation1}
N.~Austern, M.~Yahiro, and M.~Kawai,
\newblock Phys. Rev. Lett. {\bf 64}, 2649 (1989).

\bibitem{CDCC-foundation2}
N.~Austern, M.~Kawai, and M.~Yahiro,
\newblock Phys. Rev. C {\bf 53}, 314 (1996).

\bibitem{CDCC-foundation3}
A.~Deltuva, A.~Moro, E.~Cravo, F.~Nunes, and A.~Fonseca,
\newblock Phys. Rev. C {\bf 76}, 064602 (2007).

\bibitem{Tostevin2}
J.~Tostevin, F.~Nunes, and I.~Thompson,
\newblock Phys. Rev. C {\bf 63}, 024617 (2001).

\bibitem{Davids}
B.~Davids {\em et~al.},
\newblock Phys. Rev. C {\bf 63}, 065806 (2001).

\bibitem{Eikonal-CDCC}
K.~Ogata, M.~Yahiro, Y.~Iseri, T.~Matsumoto, and M.~Kamimura,
\newblock Phys. Rev. C {\bf 68}, 064609 (2003).

\bibitem{Matsumoto}
T.~Matsumoto {\em et~al.},
\newblock Phys. Rev. C {\bf 68}, 064607 (2003).

\bibitem{Egami}
T.~Egami {\em et~al.},
\newblock Phys. Rev. C {\bf 70}, 047604 (2004).

\bibitem{Matsumoto3}
T.~Matsumoto {\em et~al.},
\newblock Phys. Rev. C {\bf 70}, 061601(R (2004).

\bibitem{Matsumoto4}
T.~Matsumoto {\em et~al.},
\newblock Phys. Rev. C {\bf 73}, 051602(R) (2006).

\bibitem{LSCSM1}
T.~Egami, T.~Matsumoto, K.~Ogata, and M.~Yahiro,
\newblock Prog. Theor. Phys. {\bf 121}, 789 (2009).

\bibitem{LSCSM2}
T.~Matsumoto, T.~Egami, K.~Ogata, and M.~Yahiro,
\newblock Prog. Theor. Phys. {\bf 121}, 885 (2009).

\bibitem{THO-CDCC}
M.~Rodr\'{i}guez-Gallardo {\em et~al.},
\newblock Phys. Rev. C {\bf 77}, 064609 (2008).

\bibitem{4body-CDCC-bin}
M.~Rodr\'{i}guez-Gallardo {\em et~al.},
\newblock Phys. Rev. C {\bf 80}, 051601(R) (2009).

\bibitem{Matsumoto:2010mi}
T.~Matsumoto, K.~Kat\=o, and M.~Yahiro,
\newblock Phys. Rev. C {\bf 82}, 051602 (2010).

\bibitem{DEA}
D.~Baye, P.~Capel, and G.~Goldstein,
\newblock Phys. Rev. Lett. {\bf 95}, 082502, (2005); G. Goldstein, D. Baye, and
  P. Capel, Phys. Rev. C {\bf 73}, 024602 (2006).

\bibitem{Takechi}
M.~Takechi {\em et~al.},
\newblock Nucl. Phys. A {\bf 834}, 412c (2010).

\bibitem{Nakamura}
T.~Nakamura {\em et~al.},
\newblock Phys. Rev. Lett. {\bf 103}, 262501 (2009).

\bibitem{KMT}
A.~K. Kerman, H.~McManus, and A.~M. Thaler,
\newblock Ann. Phys. (N.Y.) {\bf 8}, 51 (1959).

\bibitem{Watson}
K.~M. Watson,
\newblock Phys. Rev. {\bf 89}, 115 (1953).

\bibitem{Minomo-DWS}
K.~Minomo {\em et~al.},
\newblock Phys. Rev. C {\bf 64}, 034602 (2011).

\bibitem{Minomo:2011bb}
K.~Minomo {\em et~al.},
\newblock arXiv:1110.3867 [nucl-th] (to be published in Phys. Rev. Lett.).

\bibitem{Sumi:2012fr}
T. Sumi {\em et~al.},
\newblock arXiv:1201.2497 [nucl-th].

\bibitem{M3Y}
G.~Bertsch, J.~Borysowicz, M.~McManus, and W.~Love,
\newblock Nucl. Phys. A {\bf 284}, 399 (1977).

\bibitem{JLM}
J.-P. Jeukenne, A.~Lejeune, and C.~Mahaux,
\newblock Phys. Rev. C {\bf 16}, 80 (1977); ibid. Phys. Rep. {\bf 25}, 83
  (1976).

\bibitem{Brieva-Rook}
F.~Brieva and J.~Rook,
\newblock Nucl. Phys. A {\bf 291}, 299 (1977); ibid. {\bf 291}, 317 (1977);
  ibid. {\bf 297}, 206 (1978).

\bibitem{Satchler-1979}
G.~R. Satchler,
\newblock Phys. Rep. {\bf 55}, 183 (1979).

\bibitem{Satchler}
G.~R. Satchler,
\newblock "Direct Nuclear Reactions", Oxford University Press  (1983).

\bibitem{CEG}
N.~Yamaguchi, S.~Nagata, and T.~Matsuda,
\newblock Prog. Theor. Phys. {\bf 70}, 459 (1983); N. Yamaguchi, S. Nagata and
  J. Michiyama, Prog. Theor. Phys. {\bf 76}, 1286 (1986).

\bibitem{Rikus-von-Geramb}
L.~Rikus, K.~Nakano, and H.~V. von Geramb,
\newblock Nucl. Phys. A {\bf 414}, 413 (1984); L. Rikus and H.V. von Geramb,
  Nucl. Phys. A {\bf 426}, 496 (1984).

\bibitem{Amos}
K.~Amos, P.~J. Dortmans, H.~V. von Geramb, S.~Karataglidis, and J.~Raynal,
\newblock in \textit{Advances in Nuclear Physics}, edited by J. W. Negele and
  E. Vogt(Plenum, New York, 2000) Vol. {\bf 25}, p. 275 (2000).

\bibitem{CEG07}
T.~Furumoto, Y.~Sakuragi, and Y.~Yamamoto,
\newblock Phys. Rev. C {\bf 78}, 044610 (2008); {\it ibid.}, C {\bf 79},
  011601(R) (2009); {\it ibid.}, C {\bf 80}, 044614 (2009).

\bibitem{Wiringa}
R.~B. Wiringa, V.~G.~J. Stoks, and R.~Schiavilla,
\newblock Phys. Rev. C {\bf 51}, 38 (1995).

\bibitem{GM}
R.~Glauber and G.~Matthiae,
\newblock Nucl. Phys. B {\bf 21}, 135 (1970).

\bibitem{DFM-standard-form}
B.~Sinha,
\newblock Phys. Rep. {\bf 20}, 1 (1975). B. Sinha and S. A. Moszkowski, Phys.
  Lett. B {\bf 81}, 289 (1979).

\bibitem{DFM-standard-form-2}
T.~Furumoto, Y.~Sakuragi, and Y.~Yamamoto,
\newblock Phys. Rev. C {\bf 82}, 044612 (2010).

\bibitem{Minomo:2009ds}
K.~Minomo, K.~Ogata, M.~Kohno, Y.~R. Shimizu, and M.~Yahiro,
\newblock J.\ Phys.\ G {\bf 37}, 085011 (2010).

\bibitem{von-Geramb}
H.~von Geramb, K.~Amos, H.~Labes, and M.~Sander,
\newblock Phys. Rev. C {\bf 58}, 1948 (1998).

\bibitem{BonnB}
R.~Machleidt, K.~Holinde, and C.~Elster,
\newblock Phys. Rep. {\bf 149}, 1 (1987).

\bibitem{Kimura}
M.~Kimura and H.~Horiuchi,
\newblock Prog. Theor. Phys. {\bf 111}, 841 (2004); M. Kimura, Phys. Rev. C
  {\bf 75}, 041302 (2007),
\newblock M. Kimura, arXiv:1105.3281 (2011) [nucl-th].

\bibitem{Gogny}
J.~Decharge and D.~Gogny,
\newblock Phys. Rev. C {\bf 21}, 1568 (1980).

\bibitem{D1S}
J.~F. Berger, M.~Girod, and D.~Gogny,
\newblock Comp. Phys. Comm. {\bf 63}, 1365 (1991).

\bibitem{WyssPriv}
R.~Wyss,
\newblock private communication (2005).

\bibitem{Ne20-sigmaI}
L.~Chulkov {\em et~al.},
\newblock Nucl. Phys. A {\bf 603}, 219 (1996).

\bibitem{Khouaja}
A.~Khouaja {\em et~al.},
\newblock Nucl. Phys. A {\bf 780}, 1 (2006).

\bibitem{30Ne}
W.~Horiuchi, Y.~Suzuki, P.~Capel, and D.~Baye,
\newblock Phys. Rev. C {\bf 81}, 024606 (2010).

\bibitem{ABC}
J.~Aguilar and J.~Combes,
\newblock Commun.~Math.~Phys., {\bf 22}, 1971, 269. E. Balslev and J.M.
  Combes,~ Commun.~Math.~Phys. {\bf 22}, 280 (1971).

\bibitem{CSM1}
S.~Aoyama, T.~Myo, K.~Kat\=o, and K.~Ikeda,
\newblock Prog. Theor. Phys. {\bf 116}, 1 (2006).

\bibitem{Aumann}
T.~Aumann {\em et~al.},
\newblock Phys. Rev. C {\bf 59}, 1252 (1999).

\bibitem{4BEIK}
D.~Baye, P.~Capel, P.~Descoubemont, and Y.~Suzuki,
\newblock Phys. Rev. C {\bf 79}, 024607 (2009).

\bibitem{Koning}
A.~J. Koning and J.~P. Delaroche,
\newblock Nucl. Phys. A {\bf 713}, 231 (2003).

\bibitem{12C-12C}
J.~Hostachy {\em et~al.},
\newblock Nucl. Phys. A {\bf 490}, 441 (1988).

\bibitem{4BDW}
S.~N. Ershov, B.~V. Danilin, and J.~S. Vaagen,
\newblock Phys. Rev. C {\bf 62}, 041001(R (2000).

\bibitem{ERT}
M.~Yahiro, K.~Ogata, and K.~Minomo,
\newblock Prog. Theor. Phys. {\bf 126}, 167 (2011).

\bibitem{Hashimoto:2011nc}
S.~Hashimoto, M.~Yahiro, K.~Ogata, K.~Minomo, and S.~Chiba,
\newblock Phys. Rev. C {\bf 83}, 054617 (2011).
%\newblock [arXiv:1104.1567 [nucl-th]].

\bibitem{Xu}
H.~M. Xu, C.~A. Gagliardi, R.~E. Tribble, A.~M. Mukhamedzhanov, and N.~K.
  Timofeyuk,
\newblock Phys. Rev. Lett. {\bf 73}, 2027 (1994).

\end{thebibliography}

\end{document}